# Predicting Foreign Exchange EURUSD direction using machine learning

Kevin Cedric Guyard
Information Science Institute, University of Geneva
Switzerland
kevin.guyard@unige.ch

Michel Deriaz
HEG, Haute Ecole de Gestion Geneve, HES-SO
Switzerland
michel.deriaz@hesge.ch

## Abstract

The Foreign Exchange market is a significant market for speculators, characterized by substantial transaction volumes and high volatility. Accurately predicting the directional movement of currency pairs is essential for formulating a sound financial investment strategy. This paper conducts a comparative analysis of various machine learning models for predicting the daily directional movement of the EUR/USD currency pair in the Foreign Exchange market. The analysis includes both decorrelated and non-decorrelated feature sets using Principal Component Analysis. Additionally, this study explores meta-estimators, which involve stacking multiple estimators as input for another estimator, aiming to achieve improved predictive performance. Ultimately, our approach yielded a prediction accuracy of 58.52% for one-day ahead forecasts, coupled with an annual return of 32.48% for the year 2022.

## 1 Introduction

The foreign exchange (Forex) market, where currencies are exchanged, is the largest in the world, with an average of $5 trillion traded daily [1]. Major participants in the Forex market include institutions, corporations, governments, and speculators. Trading in the Forex market involves exchanging pairs of currencies. Despite the high number of transactions and the attractive potential return, Forex market prediction is considered hard. Fluctuations depend on a lot of factors: economic indicators, geopolitics, human behaviors, central bank decisions, etc. For these reasons, only 2% of traders are successful in predicting Forex market movement correctly [2].

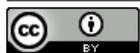

Forex can be traded using Contract for Difference (CFD). A CFD is a speculative instrument that allows a trader to bet on the rise or fall of an underlying asset. If the bet is right, the trader wins the difference between the price at the moment when the CFD is closed and the moment when it is opened. If the bet is wrong, the trader loses the difference. Thus, success in CFD trading does not hinge on forecasting price fluctuations, but rather on accurately predicting the asset's directional movement. Forecasting fluctuations is a must to determine which asset to bet on at a certain time.

In Forex trading, terms commonly used are 'bid' and 'ask' price. The bid price refers to the price a buyer is willing to pay for an asset whereas the ask price refers to the price a seller is willing to accept for an asset. The difference between the bid and ask prices is known as the spread. The spread is subject to fluctuations, partly because of variation in liquidity in the market [3].

This paper presents various methods in order to predict the daily directional movement of the Forex bid price for the EUR/USD currency pair. For clarity, the term 'direction of the pair EUR/USD' will be used throughout this paper to refer to the direction of the EUR/USD bid price. The contributions of this paper are significant and include:

- A comparative analysis of 21 machine learning algorithms, both with and without applying Principal Component Analysis (PCA) to decorrelate the input features.
- An evaluation of these 21 machine learning algorithms as meta estimators which combine the outputs of previous base models with input features for improved prediction.
- A comparative study of three dataset representations: basic features, historical features, and technical indicators.
- The incorporation of several economic indicators of the United States of America and the European Union, which influence forex traders, as input features.

This study stands out by conducting comparisons over an extended period of time (nearly 10 years) and using recent data (up to December 2022). Furthermore, the experimental setup is maintained consistent across all comparisons.

## 2 Related works

Yildirim et al. [4] suggested using a Long Short Term Memory (LSTM) model to predict the EUR/USD pair's price direction. Their study incorporated data from the EUR/USD pair, as well as from the DAX and S&P500 indices. They also included macroeconomic indicators: interest rate and inflation rate for Germany, Europe and USA. Galeshchuk and Mukherjee [5] analyzed Convolutional Neural Networks (CNNs) along with other methods to predict price movements of the EUR/USD, GBP/USD, and JPY/USD currency pairs. They found that CNN outperforms all the other methods.

Ghazali et al. [6] evaluated a Multi Layers Perceptron (MLP), a Pi-Sigma Neural Network (PSNN), a Ridge Polynomial Neural Network (RPNN) and a Dynamic Ridge Polynomial Neural Network (DRPNN) to forecast the pairs EUR/USD, GBP/USD, EUR/JPY and GBP/JPY. They found that DRPNN provides the best prediction in term of normalized mean squared error and profit. In another noteworthy study [7], Majhi et al. emphasized the importance of fast computation in financial world forecasting. They proposed two models: the Functional Link Artificial Neural Network (FLANN) and the Cascaded Functional Link Artificial Neural Network (CFLANN). They achieved good results in predicting 3 months ahead pairs GBP/USD, JPY/USD and USD/INR. In their paper [8], Perla et al. introduced a hybrid stacked AutoEncoder-based Deep Kernel based Random Vector Functional Link Network (DKRVFLN-AE) for one-day-ahead forecasting of various Forex pairs.

Additionally, there has been significant research on other stock markets that merits exploration. In [9], Di Persio and Honchar compared a MLP, a CNN and an LSTM to predict the direction and forecast the variation of the S&P500 stock market. They found that CNN provides the best results compared to both MLP and LSTM, with a mean squared error of 0.2491 in forecasting and an accuracy of 0.536 in classification. They further suggested combining various methods to achieve improved results. Fischer and Krauss proposed in [10] a comparison of LSTM with memory-free methods, such as Random Forest, Deep Neural Network (DNN) and logistic regression in order to predict the direction of the S&P500 Their conclusion was that LSTM surpassed other methods in performance. In a focused study on a particular ETF of the S&P500, Zhong and Enke [11] examined the performance DNN with and without PCA transformed dataset. They found that using a PCA transformed dataset leads to better results.

The literature review indicates that majority of research primarily utilizes historical Forex data as input features. However, both professional and individual traders who engage in forex trading also have access to historical data from other markets and economic indicators, such as the Consumer Price Index (CPI), which significantly influence their trading decisions and consequently affect Forex fluctuations. Additionally, most studies are limited to datasets spanning three to five years, often using old data from the period between 2000 and 2015. The variability in dataset periods complicates the comparison of different strategies. Furthermore, while the focus of many publications has been on neural network models, machine learning models have received less attention, except for some such as Support Vector Machine (SVM). It should be noted that most research relies on relatively small datasets—a five-year dataset typically contains approximately 1,250 daily samples. On such dataset sizes, some studies, external to finance domain, tends to show that machine learning often outperforms deep learning [12, 13].

## 3 Data

The data for this paper were gathered and compiled from various sources. They are composed of economic indicators, markets data and Forex data. The data collection spanned from April 30, 2013, to December 31, 2022. Data of year 2022 were used for evaluation while data from previous years were used for training and validation. Each data sample represents one day. Samples from days immediately before market closures were excluded from training and evaluation (since there are no activities and the price of the pair EUR/USD does not change).

### 3.1 Economic indicators

We included the following economic indicators for the United States of America (USA) and for the European Area (EA):

- The Gross Domestic Product (GDP), which represents the value of the goods produced and the services provided, is released quarterly.
- The Composite Purchasing Managers Index (Composite PMI) is obtained by summarizing monthly surveys from private sector companies and represents the economic trend of manufacturing and service of a country.
- The CPI, measuring the evolution of consumer prices, is released monthly.
- The CPI Year over Year (CPI YoY) which is the variation of the CPI from a date to the same date one year earlier. It is also known as the inflation rate.
- The interest rate fixed by central banks: these rates determine how commercial bank can borrow and lend their excess reserves. Updates are not following a special frequency.
- The Current Account Balance (CAB) which tracks the amount of money flow of a country. It is updated monthly.

Due to the fact that economic indicators are not actualized every day, we had to extend data between updates. Thus, for each sample, the value of economic indicators is the value of the last update. Even if economic indicators are related to continuous phenomena (e.g., the gross domestic product of a country is increasing continuously at every instant, not only one time at each quarter), we made the choice to have a not continuous function based only on the last update since traders who make transactions and impact the Forex are not aware of the exact value of the economic indicators at each instant but only of the update made from agencies.

Additionally, it is known that updates of economic indicators by agencies lead to increased volatility in the Forex market [14]. In this way, to capture this behavior, we also included the number of days since the last update for every economic indicator.

### 3.2 Market indices

Several market indices were included to reflect economic fluctuations in the USA and the EA:

- CAC40: index of the 40 most important French stocks.
- DAX: index of the 40 most important German stocks.
- STOXX50: index of the 50 most important Euro zone stocks.
- STOXX600: index of the 600 most important Euro zone stocks.
- DJI: index of the 30 most important New York stocks.
- NASDAQ Composite: index of more than 2500 Nasdaq stocks.
- NASDAQ100: index of the 100 most important Nasdaq non-financial stocks.
- RUSSELL2000: index of the 2000 smallest Russell 3000 stocks.
- S&P500: index of the 500 most important American stocks.

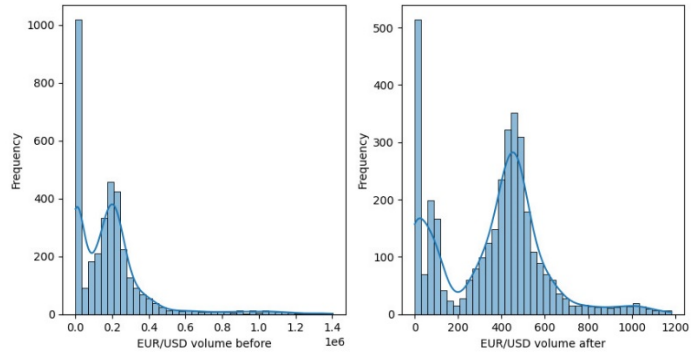

Figure 1: EUR/USD volume distribution before (left) and after (right) squared root transformation.

For each market index, the open price, close price, lowest and highest price, and daily transaction volume were included.

### 3.3 Forex pairs

The open price, close price, lowest and highest price, and daily transaction volume were included for each of the following Forex pairs: AUD/USD, EUR/AUD, EUR/CAD, EUR/CHF, EUR/GBP, EUR/JPY, EUR/NZD, EUR/USD, GBP/USD, NZD/USD, USD/CAD, USD/CHF, USD/JPY.

## 4 Preprocessing

### 4.1 Target

We define the direction of the Forex at a given day t as following:

$$Direction\ (t) = \begin{cases} 1\ if\ close\ price\ (t+1) > close\ price\ (t) \\ 0\ if\ close\ price\ (t+1) \leq close\ price\ (t) \end{cases}$$

### 4.2 Data transformation

Due to the fact that the distribution of some features was not well designed, we applied a step of data transformation. Specifically, a log transformation was applied to the European Area's CPI Year over Year (YoY), and a square root transformation was used for the central banks' interest rates and the trading volumes of market indices and Forex pairs. Figure 1 illustrates the differences in the EUR/USD trading volume distribution before and after the transformation.

### 4.3 Data representation

Our research explored three distinct data representation methods, denoted as dataset 1, dataset 2, and dataset 3. Dataset 1 was composed exclusively of daily data. Dataset 2 combined daily data with the preceding 90 days' data. Dataset 3 incorporated daily data along with technical indicators from time series of market indices and Forex pairs [15, 16]. Technical indicators are summarized in Table 1.

### 4.4 Date encoding

Dates provide valuable information when encoded effectively. Two encoding methods were proposed, depending on the prediction model used:

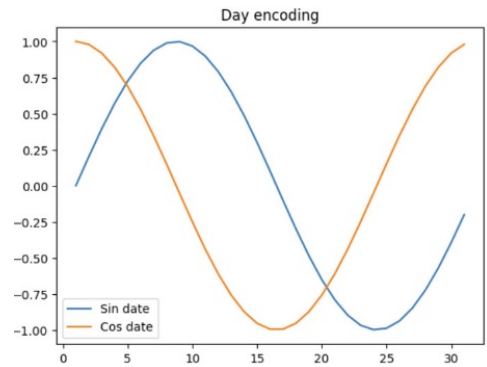

Figure 2: Day encoding using sinus and cosine for one month of 31 days.

- Tree-based models: dates were expressed as one integer for the day, one integer for the month and one for the weekday. We chose not to encode date in a one hot encoding fashion to avoid adding to many feature dimensions. Furthermore, tree-based methods are basically working well with ordinal data.
- For other models, dates were encoded using sine and cosine functions for the day, month, and weekday. Figure 2 shows an example of day encoding for one month of 31 days.

## 5 Approach

### 5.1 Models

In our study, we evaluated a variety of machine learning models, including:

- Logistic regression.
- K-Nearest Neighbors (KNN).
- Support Vector Machine (SVM) with different kernel (linear, radial basis function (RBF), sigmoid and polynomial).
- Decision tree.
- Bagging decision tree.
- Bagging logistic regression.
- Bagging KNN.

Table 1: Technical indicators derivate from Forex and market time series

| Name | Formula | Range |
| --- | --- | --- |
| Simple N days moving average (close price) | $\frac{1}{N} * \sum_{i=0}^{N-1} C(t-i)$ | $N. \in [3, 7, 14, 30, 60, 90]$ |
| Weighted N days moving average (close price) | $\frac{\sum_{i=0}^{N-1}(N-i)*C(t-i)}{\sum_{i=0}^{N-1}(N-i)}$ | $N. \in [3, 7, 14, 30, 60, 90]$ |
| Simple N days moving average (volume) | $\frac{1}{N} * \sum_{i=0}^{N-1} V(i)$ | $N. \in [3, 7, 14, 30, 60, 90]$ |
| Weighted N days moving average (volume) | $\frac{\sum_{i=0}^{N-1}(N-i)*V(t-i)}{\sum_{i=0}^{N-1}(N-i)}$ | $N. \in [3, 7, 14, 30, 60, 90]$ |
| Momentum N days | $C(t) - -C(t - -N)$ | $N. \in [1, 2, 3, 7, 14, 30, 60, 90]$ |
| Stochastic K% N days | $100 * \frac{C(t) - LL(t, t-N)}{HH(t, t-N) - LL(t, t-N)}$ | $N. \in [1, 2, 3, 7, 14, 30, 60, 90]$ |
| Stochastic D% N days | $\frac{1}{N} \sum_{i=0}^{N-1} K\%(t)$ | $N. \in [1, 2, 3, 7, 14, 30, 60, 90]$ |
| RSI N days | $100 - \frac{100}{1 + \frac{\sum_{i=0}^{N-1} \frac{Up(t-i)}{N}}{\sum_{i=0}^{N-1} \frac{Dw(t-i)}{N}}}$ | $N. \in [7, 14, 30, 60, 90]$ |
| Larry William's R% N days | $100 * \frac{H(t- N) - C(t)}{H(t- N) - L(t- N)}$ | $N. \in [1, 2, 3, 7, 14, 30, 60, 90]$ |
| A/D Oscillator | $\frac{H(t) - C(t-1)}{H(t) - L(t)}$ | |
| CCI N days | $\frac{M(t) - SM(t)}{0.015 * D(t)}$ | $N. \in [7, 14, 30, 60, 90]$ |
| ROC close N days | $100 * \frac{C(t)}{C(t-N)}$ | $N. \in [1, 2, 3, 7, 14, 30, 60, 90]$ |
| Disparity N days | $100 * \frac{C(t)}{\frac{1}{N} * \sum_{i=0}^{N-1} C(t-i)}$ | $N. \in [3, 7, 14, 30, 60, 90]$ |
| OSCP N/M days | $\frac{\sum_{i=0}^{N-1} C(t-i) - \sum_{i=0}^{M-1} C(t-i)}{\sum_{i=0}^{N-1} C(t-i)}$ | $N, M \in [(3, 7), (7, 14), (14, 30), (30, 60), (60, 90)]$ |
| MACD N days fast M days slow | $EMA_{N(t)} - EMA_{M(t)}$ | $N, M \in [(7, 21), (12, 26), (20, 34)]$ |
| MACD N days fast M days slow P days signal | $EMA_{P(MACD_{N_{M(t)}})}$ | $N, M, P \in [(7, 21, 4), (12, 26, 9), (20, 34, 17)]$ |

C(i), V(i), H(i) and L(i) are the closing price, the traded volume, the highest price and the lowest price of the day i. HH(i, j) and LL(i, j) are the highest price and the lowest price of the days between i and j. Up(i) and Dw(i) are the upward and downward price change of the day i.
$M(t) = \frac{H(t)+L(t)+C(t)}{3}$, $SM(t) = \frac{1}{N} \sum_{i=0}^{N-1} M(t-i)$, $D(t) = \frac{1}{N} \sum_{i=0}^{N-1} M(t-i) - SM(t)$, EMA is the Exponential Moving Average.

- Bagging SVM with different kernel (linear, RBF, sigmoid and polynomial).
- Random forests.
- Extra trees.
- Gradient boosting.
- Histogram gradient boosting.
- CatBoost [17].
- LightGBM [18].
- XGBoost [19].

Each model underwent analysis both with and without feature decorrelation. Feature decorrelation was accomplished using PCA [20], without reducing dimensions.

We also studied the use of meta estimators [21] (i.e. a model that used the output of other models as features to improve prediction). All the previous listed models were tested as meta estimators. Figure 3 shows an example of a meta estimator.

Prior to inputting data into the models, we also considered applying data scaling, tailored to the type of model:

- Standardization was planned as a preliminary step before feeding data into PCA (as the initial stage before the models).
- For models such as SVM, logistic regression, and their respective bagging versions, standardization was applied before data input.
- For other models, normalization was considered before feeding the model

### 5.2 Hyperparameter tuning and feature selection

We performed a step of hyperparameter tuning and feature selection using Bayesian Search [22]. For every model, 3 parallel searches are used. The first one focused solely on managing model hyperparameters. The second was able to manage the feature selection of feature categories (for example CAC40 data, DAX data etc.) in addition to model hyperparameter. The final one was tasked with conducting comprehensive feature selection (for example CAC40 low price, DAX volume) along with model hyperparameters tuning.

Table 2: Results of models without PCA.

| Model name | Dataset 1 | | | | Dataset 2 | | | | Dataset 3 | | | |
| --- | --- | --- | --- | --- | --- | --- | --- | --- | --- | --- | --- | --- |
| | Monthly | | Annually | | Monthly | | Annually | | Monthly | | Annually | |
| | Acc. | Pro. | Acc. | Pro. | Acc. | Pro. | Acc. | Pro. | Acc. | Pro. | Acc. | Pro. |
| Logistic regression | 53.7 | 9.2 | 51.4 | **29.5** | 53.3 | 13.7 | 51.1 | 8.0 | 50.4 | 2.6 | 50.4 | 6.4 |
| KNN | 54.3 | 15.5 | 49.8 | -1.5 | 48.8 | 11.2 | 53.3 | 21.9 | 50.8 | 4.8 | 52.0 | 4.3 |
| SVM linear kernel | 54.0 | **20.4** | 53.0 | 16.8 | 52.0 | 6.1 | 49.8 | 19.8 | 50.8 | 8.3 | 52.0 | 18.0 |
| SVM RBF kernel | 53.0 | -2.0 | 49.5 | -5.8 | 49.2 | -8.9 | 52.4 | 2.5 | 52.4 | -3.5 | 50.4 | -6.8 |
| SVM sigmoid kernel | 50.4 | 15.7 | 51.1 | 13.9 | 51.1 | -1.0 | 53.7 | 11.5 | 50.4 | 0.8 | 50.8 | 3.5 |
| SVM polynomial kernel | 50.1 | -1.1 | 52.0 | 4.7 | 46.6 | -15.8 | 51.7 | 10.8 | **56.2** | 21.5 | 51.1 | -4.4 |
| Decision tree | **55.9** | 14.8 | 51.4 | 7.3 | **56.5** | 5.1 | 49.2 | 0.8 | 50.1 | 3.8 | 52.0 | 10.5 |
| Bagging decision tree | 48.8 | -14.8 | 47.9 | -2.6 | 53.0 | 10.6 | 52.4 | -1.5 | 52.7 | 2.6 | 54.9 | **18.7** |
| Bagging logistic regression | 54.3 | 17.3 | 51.1 | 5.6 | 54.9 | **20.0** | 50.4 | 6.2 | 50.4 | 5.3 | 50.4 | 7.4 |
| Bagging KNN | 54.0 | 10.1 | 51.1 | 6.3 | 49.8 | 19.6 | **55.9** | **25.3** | 46.9 | -14.7 | 46.9 | -11.7 |
| Bagging SVM linear kernel | 54.3 | 19.9 | **54.0** | 20.2 | 53.7 | 15.7 | 50.8 | 6.7 | 48.5 | 3.6 | 49.8 | 5.2 |
| Bagging SVM RBF kernel | 52.4 | 12.0 | 53.3 | 26.2 | 48.2 | -12.3 | 52.0 | 7.3 | 49.2 | 7.6 | 45.3 | -8.2 |
| Bagging SVM sigmoid kernel | 49.2 | 17.0 | **54.0** | 22.2 | 48.2 | 17.2 | 51.4 | 1.9 | 50.8 | 9.4 | 49.8 | 6.3 |
| Bagging SVM polynomial kernel | 49.5 | -1.6 | 49.8 | -4.0 | 54.9 | 10.0 | 51.4 | -0.0 | 52.0 | 1.2 | 51.4 | -4.0 |
| Random forests | 50.4 | -2.5 | 50.8 | 9.6 | 49.8 | -6.7 | 48.5 | -10.0 | 50.1 | -8.1 | 51.4 | 8.2 |
| Extra trees | 48.8 | -8.1 | 48.5 | -5.7 | 52.0 | 2.7 | 54.3 | 1.1 | 50.8 | -9.2 | 48.2 | -7.0 |
| Gradient boosting | 51.1 | -6.0 | 48.8 | -7.2 | 51.7 | -13.4 | 49.5 | -5.7 | 50.1 | -3.8 | 51.1 | 0.6 |
| Histogram gradient boosting | 53.3 | 10.2 | 49.2 | 5.7 | 49.2 | 1.5 | 51.1 | 6.4 | 54.6 | -2.7 | **58.5** | 12.4 |
| CatBoost | 51.4 | -10.7 | 51.4 | 3.5 | 47.5 | -20.3 | 45.0 | -3.6 | 49.8 | -1.9 | 48.5 | -7.5 |
| LightGBM | 52.7 | 3.2 | 52.4 | 0.9 | 49.2 | -0.9 | 53.7 | 20.8 | 51.7 | 3.5 | 47.5 | -13.0 |
| XGBoost | 49.8 | 3.7 | 49.2 | -1.2 | 46.6 | -4.1 | 48.5 | 0.6 | 49.2 | -13.2 | 53.0 | 0.9 |

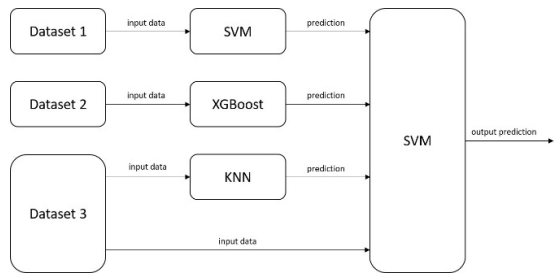

Figure 3: Example of meta estimator where a SVM (fit on dataset 1), a XGBoost (fit on dataset 2) and a KNN (fit on dataset 3) are used in first stage and then a SVM is used to predict based on the input data of dataset 3 and the prediction of the 3 models of the first stage.

Throughout the search process, information from the first search was transferred to the second, and knowledge from the second search was passed on to the third. In essence, when the first search identifies a suitable set of hyperparameters for a model that fits the entire feature set, it shares this information with the second search. Subsequently, the second search can leverage this information to enhance precision and adapt feature selection accordingly. This pattern continues between the second and third search. We adopted this approach to accelerate the convergence of the third search. Indeed, the third search has a huge search space to explore. Supplying valuable points within this complex space through a less complex search aids in enhancing the comprehension of the third search space. Figure 4 illustrates the scope of each search and the communication between them.

Hyperparameter tuning and feature selection were performed using an 8-fold rolling cross validation approach. Data from 2013/11/26 to 2019/12/31 were used only for training and data from 2020/01/01 to 2021/12/31 were used first for validation (using segments of 3 months) then added to training set before next validation segments. Slicing method is presented in Figure 5. We used accuracy as the criterion to evaluate the best set of hyperparameters and features.

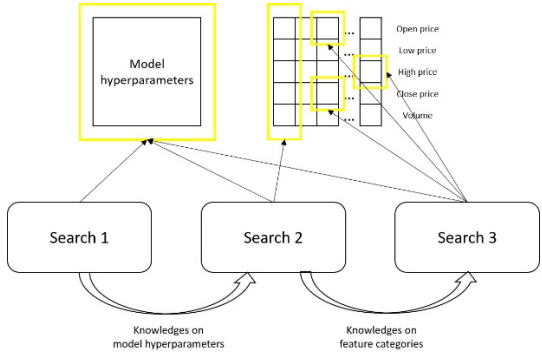

Figure 4: Bayesian searches and knowledge transfer between searches. Yellow indicates the scope of searches.

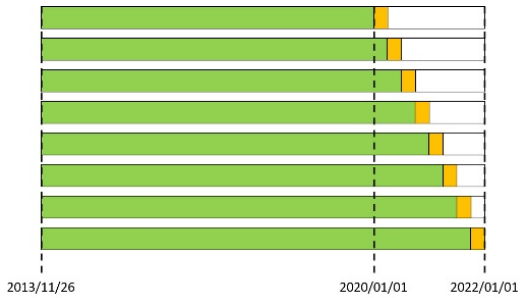

Figure 5: 8 folds rolling cross validation. Green indicates training data and orange validation data.

### 5.3 Evaluation

To evaluate the different models that we have studied, we considered two metrics: accuracy and profit. We defined profit as follows, with P(0) = 0:

$$P(t) = \prod_{i=1}^{t} 1_{y_{pred}(i)=1} * \frac{C(i)}{C(i-1)} + 1_{y_{pred}(i)=0} * \frac{C(i-1)}{C(i)}$$

We proposed two distinct approaches for model evaluation. The first one consisted of using all the training and validation data to fit models and to evaluate them on 2023 data. The second consisted of using all the training and validation data to fit models and to evaluate them on January 2023 data. Subsequently, models were retrained using the training and validation data, augmented with data from January 2023, and then evaluated on data from February 2023, and so forth.

## 6 Results

In this study, histogram gradient boosting achieved the highest prediction accuracy, scoring 58.52%. Logistic regression achieved the best profit, yielding an annual return of 29.54% (Table 2). Using PCA prior to model input resulted in lower accuracy and profit. Indeed, when using PCA, the highest achieved accuracy was 54.98%, and the greatest profit was 24.05% (Table 3). By utilizing data and predictions from basic models, meta estimators achieved an accuracy of 57.23%, slightly lower than that of the basic estimators.

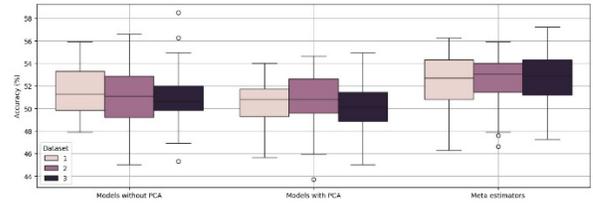

Figure 6: Distribution of the accuracy discriminated by the dataset type for models without PCA, with PCA and meta estimators.

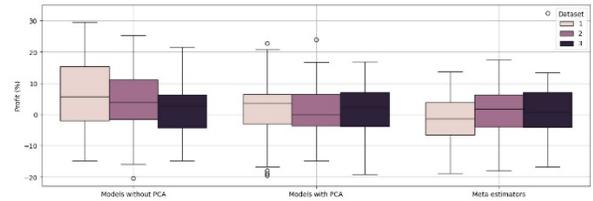

Figure 7: Distribution of the profit discriminated by the dataset type for models without PCA, with PCA and meta estimators.

However, meta estimators led to a better profit, reaching a 32.48% annual return (Table 4).

In general, meta estimators led to better accuracy, as we can observe in Figure 6. Whether for models with PCA, without PCA or meta estimators, dataset 3 reached the best accuracy. This highlights the fact that feature engineering with domain knowledge enhances the models' understanding of the problem. We can also observe that using PCA led to global worse results in term of accuracy. We hypothesize that the application of PCA prior to model input leads to a less effective representation of temporal relationships in the data.

In term of profit, basic models without PCA reached in mean the best annual returns even if they generally provided an accuracy slightly lower than meta estimators (see Figure 7). Importantly, it should be noted that accuracy and profit do not necessarily correlate. A model may succeed in more daily trades than another yet underperforms in terms of profit, succeeding in low-value trades while failing in high-value ones. Figure 8 displays a comparison of the daily trade performance of a meta estimator decision tree (accuracy = 55.31 % / profit = 32.48 %) versus a meta estimator SVM with kernel polynomial (accuracy = 55.63 % / profit = 4.25 %). As observed in the figure, the mean of the distribution of the decision tree is much higher than that for SVM, despite the SVM's higher accuracy (the SVM distribution has a bigger median but a lower mean).

The profit generated by the best-performing meta estimator, a decision tree based on dataset 3, throughout 2022 is depicted in Figure 9.

Table 3: Results of models with PCA.

| Model name | Dataset 1 | | | | Dataset 2 | | | | Dataset 3 | | | |
| --- | --- | --- | --- | --- | --- | --- | --- | --- | --- | --- | --- | --- |
| | Monthly | | Annually | | Monthly | | Annually | | Monthly | | Annually | |
| | Acc. | Pro. | Acc. | Pro. | Acc. | Pro. | Acc. | Pro. | Acc. | Pro. | Acc. | Pro. |
| Logistic regression | 51.7 | 20.7 | 50.4 | 6.2 | 51.1 | -0.8 | 50.8 | 0.6 | 52.0 | 7.5 | 48.2 | -9.4 |
| KNN | 53.3 | 8.2 | **54.0** | 7.3 | 53.3 | 13.3 | 49.8 | -0.0 | 48.5 | 0.7 | 49.2 | -5.8 |
| SVM linear kernel | 51.7 | 18.7 | 47.2 | 6.6 | 50.4 | -1.0 | 53.0 | 6.6 | 49.5 | -0.4 | 51.4 | 8.7 |
| SVM RBF kernel | 50.8 | 4.4 | 49.5 | -4.3 | 50.8 | 2.3 | 52.0 | 8.5 | 51.1 | 6.2 | 50.4 | 6.2 |
| SVM sigmoid kernel | 48.8 | 4.5 | 48.5 | -8.6 | 53.0 | -1.4 | 51.1 | 1.0 | 50.8 | -19.1 | 49.2 | 8.0 |
| SVM polynomial kernel | 51.7 | 2.2 | 51.7 | 3.4 | 52.4 | -6.2 | 50.1 | -1.3 | 49.2 | 0.1 | 49.8 | 4.6 |
| Decision tree | 46.9 | -3.6 | 49.8 | -0.9 | **54.3** | **24.0** | 49.8 | -4.2 | 50.1 | -7.0 | 48.8 | -5.4 |
| Bagging decision tree | 52.0 | 2.1 | 50.8 | -0.6 | 47.9 | -8.8 | 49.5 | -14.8 | 47.2 | -15.5 | 52.4 | 12.3 |
| Bagging logistic regression | 51.4 | 2.4 | 50.4 | 6.2 | 54.0 | 11.3 | **54.6** | 9.2 | 53.0 | 6.0 | 50.1 | -4.0 |
| Bagging KNN | 52.4 | **22.8** | 50.8 | 5.0 | 50.1 | -1.1 | 52.0 | -2.4 | 48.8 | 5.2 | 47.2 | 5.6 |
| Bagging SVM linear kernel | 49.2 | -7.5 | 50.4 | 6.2 | 49.2 | -10.4 | 54.0 | 8.2 | 51.7 | 4.9 | 50.4 | 6.4 |
| Bagging SVM RBF kernel | 51.7 | 0.7 | 51.7 | 4.4 | 49.8 | 5.8 | 51.1 | 2.9 | 50.1 | 10.9 | 48.8 | -3.5 |
| Bagging SVM sigmoid kernel | 51.7 | 16.5 | 49.5 | 5.2 | 49.2 | -1.7 | 50.4 | 5.0 | 50.1 | 1.3 | 53.0 | **16.8** |
| Bagging SVM polynomial kernel | 53.3 | 17.8 | 50.8 | 2.9 | 50.8 | -3.9 | 52.7 | -0.5 | 51.4 | 9.6 | 53.7 | 6.3 |
| Random forests | 48.2 | -16.6 | 46.6 | -17.9 | 49.2 | -0.2 | 48.5 | 0.0 | 50.1 | 2.8 | 51.1 | 7.3 |
| Extra trees | 45.6 | -19.5 | 49.5 | -11.1 | 54.3 | 16.3 | 43.7 | -13.7 | 50.8 | 0.0 | **54.9** | -3.8 |
| Gradient boosting | **53.7** | 15.4 | 51.7 | 2.8 | 49.2 | -5.6 | 46.9 | -10.9 | 52.4 | 0.9 | 45.0 | -8.1 |
| Histogram gradient boosting | 47.5 | -13.8 | 48.2 | -10.4 | 52.0 | 6.9 | 52.4 | **16.7** | 53.7 | 7.5 | 48.5 | -3.3 |
| CatBoost | 49.5 | -0.8 | 47.2 | -18.8 | 50.4 | 3.6 | 48.5 | -12.1 | 45.9 | -12.4 | 48.2 | -2.3 |
| LightGBM | 52.4 | 4.2 | 51.4 | 3.9 | 52.0 | 1.5 | 45.9 | -6.9 | 52.0 | **16.7** | 51.1 | 13.9 |
| XGBoost | 50.8 | 12.9 | 53.3 | **11.3** | 53.0 | 10.0 | 53.3 | 5.8 | 50.4 | 1.6 | 48.2 | -5.2 |

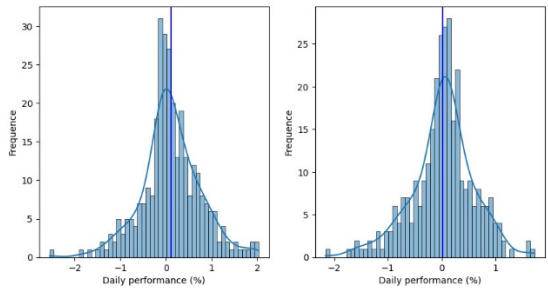

Figure 8: Comparison of the distribution of the daily return (P(t) − P(t-1)) of the meta estimator decision tree (left) and the meta estimator SVM with polynomial kernel (right), both based on dataset 3. The blue vertical line indicates the mean of the distribution.

## 7 Conclusion

This paper presents a comparison of various machine learning models, analyzed both with and without feature decorrelation employing PCA. Additionally, the study explored the use of meta estimators, which demonstrated higher accuracy than basic models. Overall, our approach yielded an accuracy of 58.52% and an annual return of 32.48% on the year 2022. It was observed that, in terms of accuracy, feature engineering informed by domain knowledge produced the most favorable results. Despite the promising outcomes obtained, we believe that machine learning approaches have the potential to yield even more impressive results. A notable limitation in predicting daily direction lies in considering economic indicators at midnight, whereas these indicators are actually released at various times during the day. Considering that the release of these indicators leads to high volatility in the Forex market, our models may not fully capture this volatility when predicting daily direction. Hence, refining predictions to an hourly or minute basis represents a promising avenue for future improvements. Another potential area for improvement would be to incorporate forecasts of economic indicators (depending on the indicator, some previsions are done with a double/triple frequency than the ground truth release).

Table 4: Results of meta estimators.

| Model name | Dataset 1 | | | | Dataset 2 | | | | Dataset 3 | | | |
| --- | --- | --- | --- | --- | --- | --- | --- | --- | --- | --- | --- | --- |
| | Monthly | | Annually | | Monthly | | Annually | | Monthly | | Annually | |
| | Acc. | Pro. | Acc. | Pro. | Acc. | Pro. | Acc. | Pro. | Acc. | Pro. | Acc. | Pro. |
| Logistic regression | 53.7 | 3.6 | 50.4 | -5.0 | 54.3 | 2.7 | 51.7 | -6.2 | 53.7 | 7.8 | 53.0 | 0.0 |
| KNN | 54.9 | 4.0 | **54.9** | -2.4 | 54.6 | 11.6 | 51.4 | -1.9 | 54.6 | 7.4 | 50.8 | -12.5 |
| SVM linear kernel | 53.7 | 7.3 | 52.4 | -8.0 | 53.0 | 6.2 | 53.0 | -2.4 | 52.0 | -0.2 | 54.0 | 7.0 |
| SVM RBF kernel | 53.3 | -0.7 | 47.5 | -14.8 | 54.9 | 12.5 | 53.0 | 1.0 | 52.0 | -0.2 | 51.4 | -2.6 |
| SVM sigmoid kernel | 52.7 | -4.2 | 51.4 | -2.9 | 53.3 | 1.5 | 47.5 | -12.1 | 54.3 | 8.4 | 51.7 | -8.4 |
| SVM polynomial kernel | 54.6 | 12.3 | 52.0 | -5.7 | 54.9 | 16.1 | 48.2 | -4.6 | 51.7 | 4.5 | **55.6** | 4.2 |
| Decision tree | 55.6 | 2.0 | 54.3 | **5.2** | 51.4 | 4.5 | **54.0** | 9.9 | 55.3 | **32.4** | 48.8 | -5.3 |
| Bagging decision tree | 54.0 | 2.3 | 50.4 | -4.0 | 53.7 | 8.1 | 53.3 | 1.1 | 53.3 | 3.0 | 52.0 | -8.8 |
| Bagging logistic regression | 53.3 | 1.5 | 50.8 | -2.0 | 53.7 | 5.8 | 53.7 | 2.1 | 54.3 | 4.7 | 54.9 | 3.1 |
| Bagging KNN | 53.7 | 7.5 | 51.1 | -9.1 | 55.3 | 11.2 | 53.0 | 6.9 | 53.3 | 5.7 | 52.7 | -0.8 |
| Bagging SVM linear kernel | 55.6 | 9.5 | 53.3 | -1.8 | 54.3 | 7.7 | 51.4 | -9.1 | 54.3 | 7.5 | 53.3 | -0.2 |
| Bagging SVM RBF kernel | 54.3 | 1.9 | 51.1 | -6.6 | 52.4 | 2.8 | 50.4 | -4.0 | 52.4 | -1.8 | 48.5 | -8.7 |
| Bagging SVM sigmoid kernel | 55.3 | **13.7** | 52.4 | 2.6 | **55.9** | **17.4** | 51.4 | 1.5 | 55.3 | 10.1 | 51.1 | 4.3 |
| Bagging SVM polynomial kernel | 55.6 | 3.4 | 49.2 | -11.4 | 54.9 | 0.7 | 52.6 | 2.9 | **57.2** | 7.1 | 54.0 | **13.5** |
| Random forests | 55.3 | 7.1 | 54.3 | 2.9 | 51.1 | -4.2 | **54.0** | -2.0 | 55.3 | 12.1 | 52.7 | 0.2 |
| Extra trees | 51.1 | -3.4 | 49.5 | -3.9 | 54.6 | 6.3 | 52.7 | 4.7 | 54.9 | 7.9 | 54.9 | 3.9 |
| Gradient boosting | **56.2** | 10.9 | 51.1 | -8.8 | 54.0 | 4.8 | 48.5 | -10.5 | 54.3 | 1.4 | 49.8 | -4.5 |
| Histogram gradient boosting | 53.3 | 7.9 | 52.7 | -0.5 | 53.3 | 7.9 | 52.7 | -0.5 | 53.3 | 7.9 | 52.7 | -0.5 |
| CatBoost | 47.9 | -8.2 | 46.3 | -18.9 | 52.7 | 5.6 | 48.2 | -10.6 | 47.9 | -0.7 | 48.2 | -9.9 |
| LightGBM | 50.8 | 10.3 | 47.2 | -9.2 | 49.2 | -3.5 | 47.9 | -10.9 | 48.8 | -12.7 | 47.2 | -16.7 |
| XGBoost | 47.2 | -18.6 | 46.9 | -14.2 | 51.4 | -5.9 | 46.6 | -17.9 | 50.1 | -5.6 | 47.9 | -12.8 |

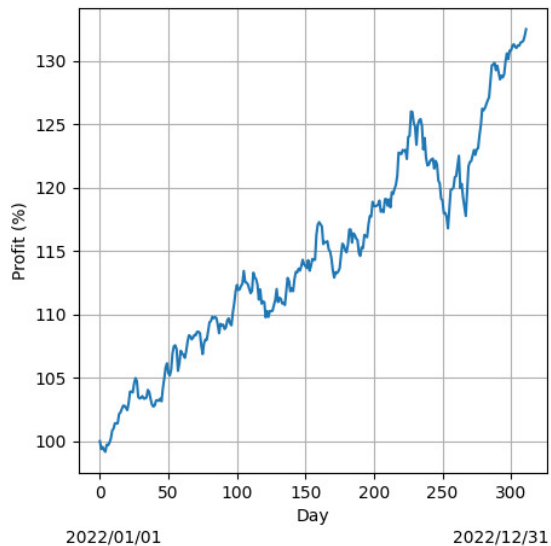

Figure 9: Evolution of the profit during year 2022 with the meta estimator decision tree based on dataset 3.

## Data sources and code

In the interests of transparency and reproducibility, we listed below the origins of the data we used:

- Economic indicators:
- https://fred.stlouisfed.org/
- https://macrovar.com/
- https://www.investing.com/
- https://data.bls.gov/
- https://data.oecd.org/
- https://www.ecb.europa.eu/
- Market data:
- https://finance.yahoo.com/
- Forex data:
- https://www.dukascopy.com/

Code is also available on demand by contacting the first author.

NB: Some of the sources require to create an account to get access to the data